\documentclass[a4paper]{VisionStyle}
\usepackage{epsfig}

\def\gtsim{\raise 2pt \hbox {$>$} \kern-1.1em \lower 4pt \hbox {$\sim$}}

\begin{document}

\title{{\it Chandra} detects inverse Compton emission from the radio
galaxy 3C 219}

\author{G.\,Brunetti\inst{1}, A.\,Comastri\inst{2},
D.\,Dallacasa\inst{3},
M.\,Bondi\inst{1}, M.\,Pedani\inst{4}, \and G.\,Setti\inst{3} } 

\institute{
  Istituto di Radioastronomia del CNR, via P.Gobetti 101, I-40129
  Bologna, Italy 
\and 
  INAF, Osservatorio Astronomico di Bologna, via Ranzani 1, I-40127
  Bologna, Italy
\and
  Dipartimento di Astronomia, Universita' di Bologna, via Ranzani 1,
  I-40127 Bologna, Italy
\and
  Centro Galileo Galilei - S/C La Palma, 38700 TF Spain}

\maketitle 

\begin{abstract}

We report the results from a {\it Chandra} observation of the powerful
nearby (z=0.1744) radio galaxy 3C 219. We find evidence for non--thermal
X--ray emission from the radio lobes which fits fairly well with a 
combination of IC scattering of CMB and of nuclear photons.
The comparison between radio synchrotron and IC fluxes yields
a magnetic field strength significantly lower ($\sim 3$) than
that calculated under minimum energy conditions; the majority 
of the energetics is then associated to the relativistic particles.

\keywords{Radiation mechanisms: non-thermal -- Galaxies: active --
Galaxies: individual: 3C 219 -- Radio continuum: galaxies --
X-rays: galaxies}
\end{abstract}

\section{Introduction}

A full understanding of the energetics and energy
distribution of relativistic particles in jets and lobes
of radio galaxies and quasars is of basic importance for
a complete description of the physics and time evolution
of these sources.
It is assumed that the relativistic electrons
originated in the nuclear regions of powerful radio sources,  
are channeled into the
jet and re--accelerated in the radio hot spots,
which mark the location of strong planar shocks formed at
the beam head of a supersonic jet
(e.g. Begelman, Blandford, Rees, 1984) and then diffuse in the
radio lobes.

Extended non--thermal X--ray emission from
the lobes of radio galaxies and quasars
is produced by IC
scattering of Cosmic Microwave Background (CMB) 
photons (e.g. Harris \& Grindlay 1979),
and/or nuclear photons (Brunetti, Setti \& Comastri 1997).
In the first case one is sampling the relativistic
electrons with Lorentz factor $\gamma$\gtsim$10^3$, while 
the X--rays from IC scattering of
the nuclear far--IR/optical photons are mainly powered
by $\gamma=100-300$ electrons
whose synchrotron emission typically falls
in the undetected hundred kHz frequency range.
As a consequence the study of the diffuse X--ray emission
from the radio lobes 
is a unique tool to constrain the spectrum of the relativistic 
electrons 
and hopefully the acceleration mechanisms at work.

Non--thermal 
X--ray emission from the radio lobes has been discovered by {\tt ROSAT} and
{\tt ASCA} in a few nearby radio galaxies, namely Fornax A (Feigelson et
al. 1995;
Kaneda et al. 1995; Tashiro et al. 2001), 
Cen B (Tashiro et al. 1998), 3C 219 (Brunetti et al. 1999)
and NGC 612 (Tashiro et al. 2000).
By combining X--ray, as IC scattering of CMB photons, 
and synchrotron radio flux densities
it has been possible to derive magnetic field strengths
lower, but within a factor $\sim 3$ of the
equipartition fields.
In general, these observations have been complicated due to 
the weak X--ray brightness, relatively low count statistics and 
insufficient angular resolution of the instruments.

\begin{figure*}[ht]
\begin{center}
\epsfig{file=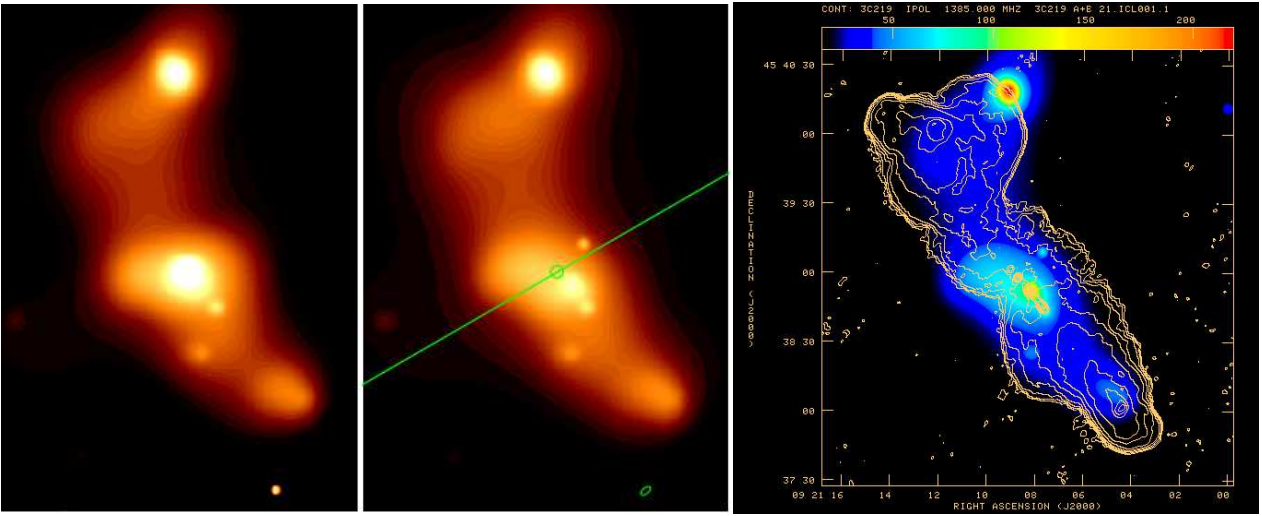, width=17cm}
\end{center}
\caption{{\bf Panel a)}: 0.3--7 keV smoothed {\it Chandra} 
image; {\bf Panel b)}: 0.3--7 keV image smoothed after subtraction of
the nucleus and of the pile up streaks; the intensity
scales in panel a) and b) are slightly different.
In the subtracted image are also visible the X--ray counterparts
of the inner knot of the radio jet (panel c) 
and of the nucleus of the radio 
galaxy baby-3C 219 (revealed in the high resolution radio
images, e.g. Perley et al. 1980).
The position of the nucleus of 3C 219 and the pile up direction
are reported in the panel; 
{\bf Panel c)}: VLA 1.4 GHz radio contours superimposed
on the {\it Chandra} image of the panel b).}
\end{figure*}

The enhanced capabilities of the
{\it Chandra} X--ray observatory
make now possible to image (on arcsec scale)
the radio lobes of powerful radio galaxies and quasars
and to disentangle the non--thermal emission from the
thermal component (if any) 
and from the nuclear component.
Non--thermal emission the from radio lobes of 
relatively compact and powerful objects
has been successfully detected with 
{\it Chandra} in the case of the radio galaxy
3C 295 (Brunetti et al., 2001) and, more recently,
in the counter lobes of the radio loud quasars 
3C 207 (Brunetti et al. 2002) and 3C 179 (Sambruna et al. 2002).
These observations are well interpreted as IC scattering of 
IR photons from the corresponding nuclear source,  thus
providing a clear evidence for the presence of 
low energy electrons ($\gamma \sim 100$) in the radio volumes.

In this paper we report on the {\it Chandra} observation of
the nearby powerful radio galaxy 3C 219.
Combined {\tt ROSAT PSPC}, {\tt HRI} and {\tt ASCA} observations have
previously found 
evidence for extended non--thermal emission.
This emission has been interpreted as IC
scattering of CMB and of IR
nuclear photons with the resulting averaged
magnetic field intensity in the radio lobes being a factor of $\sim 3$
below the equipartition value (Brunetti et al. 1999).
However, the presence of a bright point--like 
nuclear source, smearing the
diffuse emission in the {\tt ROSAT HRI} image within $\sim 15$ arcsec 
distance from the nucleus, and the impossibility to perform spatially 
resolved spectroscopy with the past satellites has required  
a follow up observation with {\it Chandra} to better
image the extended
emission and to confirm its non--thermal origin.
We show that the {\it Chandra} observation 
essentially confirms the previous findings.
A more detailed discussion of the results will be
presented in a forthcoming paper.
$H_0=50$ km/s/Mpc and $q_0=0.5$ are used throughout.

\section{Target and data analysis}

3C 219 is a nearby (z=0.1744) powerful radio galaxy identified
with a cD galaxy
of magnitude M$_V$=-21.4 (Taylor et al.1996), belonging to a non Abell
cluster (Burbidge \& Crowne 1979).
The radio structure is well studied (Perley et al. 1980,
Bridle et al. 1986, Clarke et al. 1992): it is a classical double--lobed
FRII radio galaxy
that spreads over $\sim$ 180\arcsec on the sky plane corresponding to a
projected size of $\sim$ 690 kpc.

In order to improve the available radio information, to
image the radio lobes down to very low brightness and 
to derive radio spectral index maps,  
we obtained and analyzed 12 hrs of new observations with the VLA
at 1.4 and 5 GHz in C and D configuration and combined
them with the observations available in the archive;
a deep 1.4 GHz VLA image is reported in Fig.1c
(Dallacasa et al., in preparation).

\subsection{Chandra Data}

The target was observed with the {\it Chandra} X--ray
observatory on 2000 October 11th for about 20 ksec.
Although only half of the chip was used in the window mode 
the nuclear source is significantly affected by pileup 
and is not considered any further.
The raw level 1 data were re--processed using the latest
version (CIAO2.2) of the CXCDS software.
The target was placed about 40'' from the nominal aimpoint
of the  back illuminated ACIS S3 chip.
There is evidence of an increased background count rate towards 
the end of the observation.
The corresponding time intervals  
were filtered out leaving about 16.8 ksec of useful data
which were used in the spectral and imaging
analysis described below.
Given that we are interested in the study of the origin of the faint
diffuse emission from the radio lobes the background subtraction 
is an important issue. 
The particle background can be reduced significantly 
in the observations carried out in the Very Faint Mode 
(Vikhlinin 2001). We have employed this technique
with the recommended choice of parameters. As a result the
quiescent background is reduced up to 30\%.
Fig.1a--c show several features:
relatively faint diffuse emission coincident with the radio lobes
showing a brightness increment in the innermost part of the
northern lobe
and in the region of the back flow of the southern hot spot,
a bright clump at north west on the boundary of the northern
radio lobe, 
and two distinct knots at 10--30 arcsec south of the nucleus.
These X--ray knots are spatially coincident with
the two radio knots of the main jet; the knots and the nuclear 
properties are not discussed in this contribution.

X--ray spectra have been extracted
using appropriate response and effective area functions taking into
account the source extension and weighting the detector response 
and effective area according to the source spectrum.

\subsection{The background cluster}

The spectrum of the north west clump is well fitted by a
thermal model with a relatively low temperature, $kT \sim 2$ keV
(Fig.2).
The derived absorbing column density
(with a best fit value of $N_H \sim 10^{21}$cm$^{-2}$) is
marginally consistent
at the 1\% confidence level with the Galactic value
but we note that the count statistics is rather poor.

A R-band HST observation taken from the archive shows the presence of
an excess of optical galaxies at the position of the north west clump,
possibly indicating the presence of a cluster/group.
We have obtained the spectrum of the most luminous galaxy of
the group with the 3.5mt. Telescopio Nazionale Galileo at La Palma
(TNG) and identify
it with an elliptical galaxy at a redshift z=0.389.
If we assume that this is the CD galaxy of the group/cluster
which emits the observed X--rays, the
derived 0.1--10 keV luminosity is $\sim 2 \times 10^{43}$erg/s which
is only a factor of $\sim 2$ below the best fit luminosity expected
from a 2 keV temperature cluster by adopting the
Luminosity--Temperature relationship of Arnaud \& Evrard (1999).
Thus we identify the X--ray
clump as emission from a background cluster/group at $z=0.389$.

\begin{figure}[ht]
\begin{center}
\epsfig{file=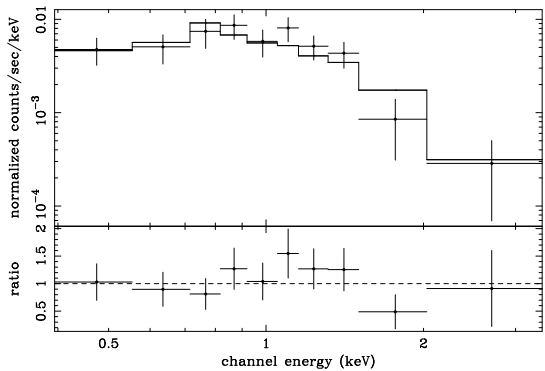, width=8cm}
\end{center}
\caption{ The 0.4--3 keV spectrum extracted from the cluster region
is shown compared with a thermal model with kT=2 keV, 0.3 solar
abundances and $N_{\rm H}=8\times 10^{20}$cm$^{-2}$.}
\end{figure}

\subsection{The non--thermal emission}

The morphology and scale of the
faint diffuse X--ray emission is very similar to that of the radio
lobes thus suggesting a non--thermal origin.
X--ray emission from IC scattering of nuclear photons would 
outweigh that from IC scattering of CMB at a distance from the 
nucleus : 

\begin{equation}
R_{\rm kpc} <
70 \times L_{46} 
 \left( 1 - \mu \right)^2 \left( 1 + z \right)^{-2}
 \label{conf}
\end{equation}

where $L_{46}$ is the isotropic
nuclear luminosity in units of $10^{46}$erg/s in the far--IR to optical
band and $\mu$ is the cosine of the
angle between the direction of the nuclear seed 
photons and the scattered photons (roughly speaking 
the angle between radio axis and
line of sight with $\mu$ negative for the far lobe and positive
for the near one).

Since we detect diffuse  X--ray emission coincident with the
radio lobes out to a distance of $\sim$300 kpc from the nucleus
with a roughly constant brightness (Fig.1),  
we conclude that the IC scattering of CMB photons is the 
process which accounts for the majority of the observed X--rays.
On the other hand, we notice the presence of a net increment of the 
X--ray brightness in the innermost $\sim 18$ arcsec 
(projected size $\sim$70 kpc) of the northern
lobe (i.e., the far/counter lobe).
If this is of IC origin, as suggested by the fact that 
the X--rays are strictly related to the morphology of the radio
contours (Fig.1c), it should require an
additional source of seed photons, most likely from the nucleus
(under the reasonable assumption that the density of
the relativistic electrons does not significantly increase inward).
By imposing an equal contribution to the X--ray flux
from the IC scattering 
of CMB and nuclear photons at a projected distance of $\sim 18$ arcsec 
in the northern lobe, and by 
assuming an angle between the radio axis and the plane of the sky of 
20--30$^o$, Eq.(\ref{conf}) yields a far--IR to optical luminosity of
the hidden quasar of $\sim 5 \times 10^{45}$erg/s 
which, indeed, is only a factor $\sim 2$ lower  
than that independently estimated by Brunetti et al.(1999) 
making use of a quasar SED normalized to the nuclear X--ray flux.
We stress that the present estimate does 
not depend on the energy densities of the electrons and of the
magnetic field in the radio lobes. 

\begin{figure}[ht]
\begin{center}
\epsfig{file=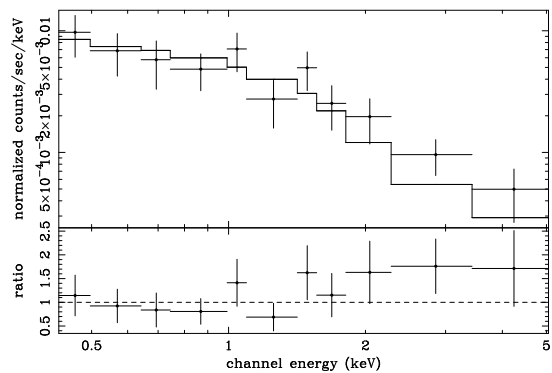, width=7.5cm}
\end{center}
\caption{The 0.4--5 keV spectrum extracted from the southern lobe
is shown compared with a power law model with the same slope ($\alpha=0.7$) 
of the radio spectrum. Despite the relatively poor statistics the
excess of the data with respect to the model at $\geq 2$keV
may indicate that the 
X--ray spectrum is slightly flatter than the radio one.}
\end{figure}

We performed spectral analysis of the diffuse emission extracting the
spectrum from the southern radio lobe which has a better statistics
(i.e. the highest ratio between source and background counts).
A power law, with the absorption fixed at the Galactic
value, does provide an acceptable description of the data (Fig.3) with
the derived X--ray spectral index in the range $0.3\pm0.4$
flatter, but consistent (90\% confidence level), with the 
spectral index derived from the integrated synchrotron spectrum of 
the radio galaxy ($\alpha = 0.7-0.8$; e.g., Laing et al. 1983).

A slightly worse, but still statistically good description of the data,
is obtained using a thermal model with abundances fixed at 0.3 solar.
It is important to point out that although the quality of the data
is not such to discriminate between the two options
(power--law and thermal)
the derived best fit temperature is 
$\sim$20 keV (${\rm kT} > 5.5$ keV at 90\% conf. level for one relevant
parameter)
and thus it appears to be in contrast with the relatively 
low X--ray luminosity of the diffuse emission.
Indeed, from the luminosity--temperature correlation
(e.g., Arnaud \& Evrard 1999) one has that
thermal emission from a hot cluster (with ${\rm kT} > 6$ keV)
would provide a luminosity $\geq 10^{45}$erg s$^{-1}$,
whereas we find that the 0.5--8 keV luminosity 
of the southern lobe is only $\sim 10^{43}$erg s$^{-1}$, 
more than about two orders of magnitude lower than 
that of a cluster with the same temperature.
Due to the large area covered by the faint diffuse emission, 
we have carefully
evaluated the influence of background subtraction
on the derived spectral parameters using different extraction regions
over the detector. In all the cases the 
lower limit on the thermal model temperature turns out 
to be of about 6 keV. 

By assuming a power law spectral index $\alpha =0.7$, 
consistent with both the X--ray and radio spectra, 
from the measured
synchrotron and IC fluxes extracted from the southern lobe,  
we derive a magnetic field $B \simeq 2.8 \mu$G which is a factor of 3 below
the minimum energy field derived with equipartition formulae
based on a low energy cut--off in the electron spectrum  
(Brunetti et al. 1997, assuming $\gamma_{\rm low}=50$) or a factor of
2.2 below the equipartition field calculated with the standard  
equipartition formulae (e.g., Packolczyk 1970).
As a consequence, the electron energy density in the southern
radio lobe is $\sim 60$ times that of the average magnetic field.
A similar result (but of course less robust) 
was previously obtained from the combined {\tt ROSAT} and {\tt ASCA} 
study of 3C 219 (Brunetti et al. 1999).

\section{Conclusion}

The relatively short (16.8 ksec) {\it Chandra} observation of the radio 
galaxy 3C 219 has successfully detected diffuse X--ray emission 
from the radio lobes and X--ray emission associated to the
main radio jet.
The morphology, extension and spectrum of the diffuse 
emission strongly support a non--thermal origin, most likely 
IC scattering of CMB photons.
In addition the net brightness increment observed in the innermost
$\sim 70$ kpc of the northern radio lobe (the far lobe) 
suggests an additional contribution from IC scattering of nuclear
photons for a reasonable far--IR to optical isotropic luminosity of
the hidden quasar of $\sim 5 \times 10^{45}$erg/s.

By comparing the radio synchrotron and IC/CMB 
X--ray fluxes from the southern radio lobe
we obtain a magnetic field strength approximately 
a factor $\sim 3$ lower than the equipartition value.

Additional data from a deeper {\it Chandra} observation would allow 
to perform detailed, spatially resolved X--ray spectroscopy of the 
diffuse emission and to compute $B$-field intensity map/structure 
in the radio lobes if combined with the available radio data.

Finally, the relatively bright X--ray clump at north west on the border 
of the northern radio lobe (Fig.1) is well fitted by thermal emission 
from a relatively cold plasma ($\sim 2$ keV) in a cluster/group of 
galaxies at a redshift $z=0.389$.

\end{document}